\documentclass[english,letterpaper,twocolumn,showpacs,pra,aps]{revtex4}
\usepackage{graphicx}
\usepackage{amssymb}

\makeatletter

\large  

\input epsf

\makeatother

\usepackage{babel}
\makeatother
\begin{document}

\title{Interacting bosons in a nearly-resonant potential well}

\author{Joseph P. Straley $^{(1)}$ and Eugene B. Kolomeisky $^{(2)}$}

\affiliation{$^{(1)}$ Department of Physics and Astronomy, University of Kentucky,
Lexington, Kentucky 40506\\
$^{(2)}$ Department of Physics, University of Virginia, 382 McCormick
Rd., P. O. Box 400714, Charlottesville, Virginia 22904-4714}

\begin{abstract}
We establish that the ability of a localized trapping
potential to bind
weakly-interacting bosons is dramatically enhanced in the vicinity of
the threshold of formation of the single-particle bound-state of the trap.
Specifically, for repulsive particles and a super-threshold trapping
potential the 
equilibrium number of bound bosons and the size of the  ground state diverge upon approaching the
single-particle threshold from above.  For attractive interactions and
a sub-threshold trap a collective bound state always 
forms for a sufficiently large number of bosons despite the
inability of interparticle attraction alone to form a two-body bound
state. 
\end{abstract}


\maketitle

\section{Resonant binding}

It is empirically known that a surprisingly large number of
short-range interactions operating in Nature have resonant character:  
the
two-particle attraction
is either barely sufficient to form a two-body bound state or misses the
binding threshold by a small amount.   In nuclear physics the
prominent examples include the neutron-nucleon and  neutron-alpha particle 
attractions which are nearly binding \cite{Richard}.   In atomic
physics the same is true for interactions between the atoms of
$^{3}He$ \cite{Aziz} and of those of the spin-polarized Hydrogen family
\cite{Hydrogen}.  At the same time two $^{4}He$ atoms form a
weakly-bound dimer whose binding energy is four orders of
magnitude smaller than the depth of the $^{4}He$-$^{4}He$ potential
well \cite{He}.  

The proximity to the two-body binding
threshold can be quantified by the $s$-wave scattering length whose 
magnitude significantly exceeds the range of interparticle
forces, thus implying an effectively long-range interaction
\cite{Richard}.  This regime is fundamentally interesting because systems 
consisting of resonantly-interacting particles may exhibit a series of
effects which are independent of the microscopic details of the 
interaction between the constituents.  One such 
phenomenon is the well-understood Efimov
effect \cite{Efimov},  which occurs in a system of three bosons where
resonant two-body forces trigger a formation of an arbitrarily large
number of loosely bound levels in a three-particle
system.   Conclusive experimental observation of the Efimov physics became possible 
only
recently in an ultra-cold gas of $Cs$ atoms \cite{Grimm} because, in
contrast to other nearly-resonant systems found in Nature, the two-body
scattering length and thus interparticle
interactions in a cold gas can be precisely controlled by using Feshbach
resonances \cite{Feshbach}.  This allowed the measurement of the dependence of the three-body recombination rate on the two-body scattering length which contains signatures specific to the Efimov effect \cite{Braaten}. 

The goal of this paper is to establish the existence of a novel resonant
effect which may also be observed in a cold bosonic gas:  we will
demonstrate that the number of particles bound by a nearly-resonant
potential well increases as the well is made more shallow, diverging
exactly at the single-particle threshold.  The fact that such an
effect might exists is implied by variational analysis given in our
previous work \cite{KSK}.  Similar to the Efimov
effect the ultimate origin of this counterintuitive behavior is an
effectively large attraction range of the potential well.  Whereas the 
Efimov effect requires an attractive part of the physical potential to 
bind three particles, the phenomenon discussed below takes place in
the presence of interparticle repulsion.

In the laboratory attractive wells have
been realized using the optical dipole force \cite{Grimm00}, 
in which atoms are attracted to the intense region of
a focused laser beam.  Strong confinement in three dimensions can be obtained 
at the intersection of two beams, each typically having diameter of
about 15 $\mu$m \cite{Adams95}.  The binding properties of such traps can be
tuned by adjusting the laser power.    Even tighter traps can be
achieved by using holographic techniques \cite{Newell02},
optical superlattices \cite{Wasik97}, 
near-field \cite{Shin03}, and white-light techniques
\cite{sackett03}.  

Additionally, tuning the parameters of such localized potential placed
next to the classical edge of a Bose-Einstein condensate, might allow 
manipulation of a well-defined number of particles \cite{KSK};  it also 
provides a new context for studying the quantum many-body problem,
as this involves the states of an interacting boson system.  The 
experimental feasibility of such a setup was discussed in
Ref. \cite{KSK} as well as reasons why tighly focused potential is needed.

\section{The model}

We consider $n$ interacting bosons of mass $m$ in the
presence of a well-localized attractive potential $U(\textbf{r})$.
We assume a system of sufficiently low density so that the range of 
interactions between particles need not be included as a parameter.  Therefore
we study the Gross-Pitaevskii (GP) energy functional \cite{GP}
\begin{equation}
\label{GPE}
E=n \int d^{3}x \left ( {\hbar ^{2}\over 2m} (\nabla \psi)^{2}
+ U({\bf r}) \psi^{2} + {\frac {g(n - 1)}{2}} \psi ^{4} \right )
\end{equation}
subject to the normalization condition
\begin{equation}
\label{norm}
\int \psi^{2} d^{3}x = 1 
\end{equation}
Here we have assumed that the many-body wavefunction can be written as
a product of the single-particle functions $\psi({\bf r})$.   The parameter $g$ represents
the short-ranged interaction between the particles.   The external
potential $U(\textbf{r})$ 
will be chosen in the form of an attractive "shell" of radius $a$:
\begin{equation}
U({\bf r}) = - a V \delta (r-a)
\end{equation}
because this is a simple
form that will bind particles without having singular behavior in the wavefunction at
the origin.  
The case of more realistic potentials will be discussed in Section VI.
For the case of one particle, there is a single bound state for 
$V > V_{c} = \hbar^2/2ma^{2}$, and this condition will
continue to be relevant for the many-body problem.

\section{results from an approximate wavefunction}

The ground-state wavefunction minimizes the energy functional
(\ref{GPE}).  We can deduce its general form from inspection.
If the ground state is to be bound, it will have to be relatively large in the
well-localized region where $U(\textbf{r})$ is nonzero.
At sufficiently large distances it is decreasing more rapidly than 
$1/r$, to ensure normalizability.  
It will pass from large to small in a way that keeps the gradient term
small; this suggests $\psi \approx 1/r$ at intermediate distances.
The effect of the interparticle repulsion
(the last term in Eq. (\ref{GPE})) will be to suppress 
large values of $\psi$.
 
In a previous paper \cite{KSK} these considerations led us to study the two-parameter trial wavefunction
such that
$\psi = C$ for $r < a$, and
$\psi(r) = C a \exp(-\alpha r + \alpha a)/r$ for larger $r$. 
We will briefly recapitulate the results.
The form of the wavefunction for larger $r$ is the exact wavefunction for a bound state of
noninteracting particles, and will remain a good approximation to 
the interacting wavefunction at large $r$, where $\psi^{4}$ will be
negligibly small.
The wavefunction extends to a distance of order $1/\alpha$, which will
be large for large $n$ and also for $V$ close to $V_{c}$.  Thus we
are particularly interested in the case of small $\alpha$.
Integration of Eq. 
(\ref{norm}) gives
\begin{equation}
\label{norm2}
4 \pi (\frac {1}{2 \alpha } + \frac {a}{3}) a^2 C^{2} = 1
\end{equation}  
We note that for small $\alpha$ this reduces to $C^{2} \approx
\alpha /a^{2}$,
rather than $\alpha^{3}$ (i.e. the reciprocal of the "volume" of
the wavefunction).  This implies that the two length scales $a$
and $1/\alpha$ both play roles; the limit $a \rightarrow 0$ cannot be
taken.
Eq. (\ref{GPE})  
leads to
\begin{eqnarray}
\label{varEn}
{\frac {E}{4 \pi n a^3}} = (V_{c} - V + {\frac {1}{2}} \alpha a V_{c})  C^{2}
+
 {\frac {2}{3}}g(n-1)C^4 
\nonumber
\\
- 2g(n-1)\alpha a e^{4 \alpha a} E_{1} (4 \alpha a ) C^{4}
\end{eqnarray}
where $E_1(x)$ is the exponential integral \cite{AS}.  
 
Using Eq. (\ref{norm2}) to eliminate mention of $\alpha$, the leading
terms of Eq. (\ref{varEn}) are
\begin{eqnarray}
\label{varEnApp}
{\frac {E}{4 \pi n a^3}} = (V_{c} -V)  C^{2} + \pi (a^3 V_{c} +  {\frac {2}{3 \pi}} g (n-1) ) C^{4}
\end{eqnarray}
Minimizing this expression with respect to $C$ gives
an estimate for the energy.
Assuming $g > 0 $ (i.e.   the particles repel
each other),   gives for $V \ge V_{c}$
\begin{equation}
E= - {{ n  (V - V_{c})^{2}} \over {V_{c} + 2g(n-1)/3\pi a^3}}
\end{equation}
For large $n$ this becomes independent of $n$, and the corresponding value 
of $\alpha$ is small.  
The energy at $n \rightarrow
\infty$ is greater or less than the energy at $n = 1$, depending on whether
$g$ is greater or less than $g_c = 3 \pi a^{3} V_{c}/2$.  For $g$ less than
$g_c$, this would seem to indicate that the potential can bind an 
infinite number of particles (with a very swollen wavefunction); however,
including the last
term of (\ref{varEn}) gives a shallow minimum for $g < g_{c}$, and then for large $n$ the state is unstable against 
the unbinding of particles (i.e. it is within the continuum of a
state with smaller $n$).
Our main goal is to
establish whether this minimum is real, and if this is the case, what
is its behavior as the single-particle
threshold is approached from above, $V \rightarrow V_{c} + 0$.

The two regimes for the interparticle strength (the cases $g > g_{c}$ and $g < g_{c}$)
are in essence a small-$n$ effect.  The ground-state energy for two particles is greater
or lesser than the energy for one, depending on whether the interaction parameter is large
or small.
  
This shows that for $V > V_{c}$ there is a $n$-body bound state; however, the total
energy for large $n$
becomes independent of $n$, so that the chemical potential decreases to zero.  This 
occurs because for small $\alpha$ the single-particle wavefunctions extend to
large distances from the attractive center, making it correspondingly unlikely that two particles will be
at the same place, so that the interparticle interactions can be
effectively made as small
as one likes (by tuning $\alpha$).  The effect depends on the way the
size of the ground state scales
with the number of bound bosons, and does not occur in lower
dimensions \cite{KSK}.

\section{Gross-Pitaevskii equation}

The approach just outlined has
the advantage that the mathematics is transparent, and yields results for
the energy and wavefunction in which the role of the various parameters in the
problem is explicit.   It is a robust method that gives an upper bound on the exact
ground-state energy.   However, the results are only as good as the trial wavefunction.
Therefore we will find the true minimum of Eq. (\ref{GPE}) by
numerical means.  This gives the best possible wavefunction that
can be written as a simple product of single particle wavefunctions.
In the process it verifies that the simple variational function
is qualitatively correct, and puts the theory of resonant
binding on solid footing.  Additionally we will show
that non-trivial cooperative bound states also form for a sub-critical
well in the presence of a weak interparticle attraction. 

\begin{figure}
\includegraphics[
  width=1.0\columnwidth,
  keepaspectratio]{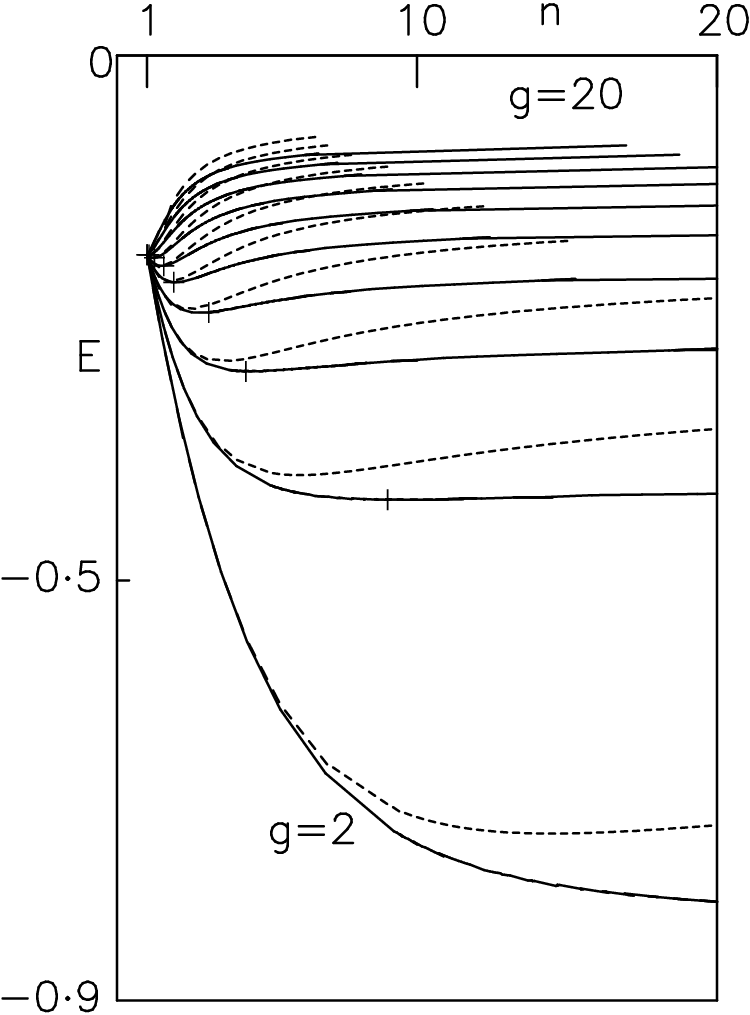}
\caption{Dependence of the ground-state energy on the number of
particles $n$, for super-threshold well, $V = 1.5$, and various values of the interaction parameter $g = 2,
4, 6, ...20$ (bottom to top).  The solid line is the numerical solution
to Eq. (\ref{diffGPE}) substituted into Eq. (\ref{relation}); 
the dotted line is the minimum value of Eq. (\ref{varEn}).  Dimensionless variables defined in the main text are used. }
\end{figure}
In what follows we will choose $a = 1$ and $V_{c} = \hbar^2/2ma^{2} = 1$,
which sets the
length and energy scales for the problem.   

Taking a variational derivative of Eq.(\ref{GPE}) gives the GP 
equation \cite{GP}
\begin{equation}
\label{diffGPE}
- {1 \over r} {d^{2} \over dr^{2}} (r \psi) - V \delta (r-1) \psi + \gamma \psi^{3} = \epsilon \psi
\end{equation}
where $\gamma = g(n-1)$.  Unlike the case of the linear Schr{\"o}dinger equation, the eigenvalue $\epsilon$ for
which the GP  equation (\ref{diffGPE}) has a normalizable
solution is not directly proportional to the minimum value of Eq. (\ref{GPE});
however, they are related by
\begin{equation}
\label{relation}
E = n \epsilon - \frac {1}{2} g n(n-1) B,
\end{equation}  
\begin{equation}
\label{B}
B = \int \limits_{0}^{\infty} \psi^{4} 4\pi r^{2} dr
\end{equation}

We solved Eq. (\ref{diffGPE}) by integrating from $r=0$ with initial conditions
$\psi = 1$, $d \psi/dr = 0$  and chosen values for  $V$,  
$\gamma$, and $\epsilon$.
The integration was in the form of a 
power-law representation of $\psi(r)$ for $r < 1$, which avoided the problems
coming from the singular form of the GP equation at small $r$ and gave accurate
values for $\psi$ and its derivative at $r=1$;  thereafter, we used second-order
Runge-Kutta integration with equal-sized steps in the variable $y=\sqrt r$.
In general this gave a function that diverged to plus or minus infinity
at large $r$.   By tuning $\epsilon$ we could find a case for which $\psi$ remained
small to chosen large values of $r$.   This condition defines the eigenvalue 
$\epsilon$ to high accuracy.
Once a normalizable solution had been found, it was rescaled to satisfy the normalization
condition (\ref{norm}).  This also changes the value of $\gamma$, but
not $V$ or $\epsilon$; only at the end of the calculation do we find
what value of the particle number $n$ the solution 
corresponds to.
 
Figure 1 shows how the ground-state energy varies with the number of
bosons $n$,
and compares the numerical solution of the GP equation, 
Eqs. (\ref{diffGPE})-(\ref{relation}), to the
variational result (the minimum value of Eq. (\ref{varEn})).
The parameter $V = 1.5$, and $g = 2, 4, 6,..., 20$.  
The GP solution is always below the variational prediction, 
because it uses a better wavefunction.

\begin{figure}
\includegraphics[
  width=1.0\columnwidth,
  keepaspectratio]{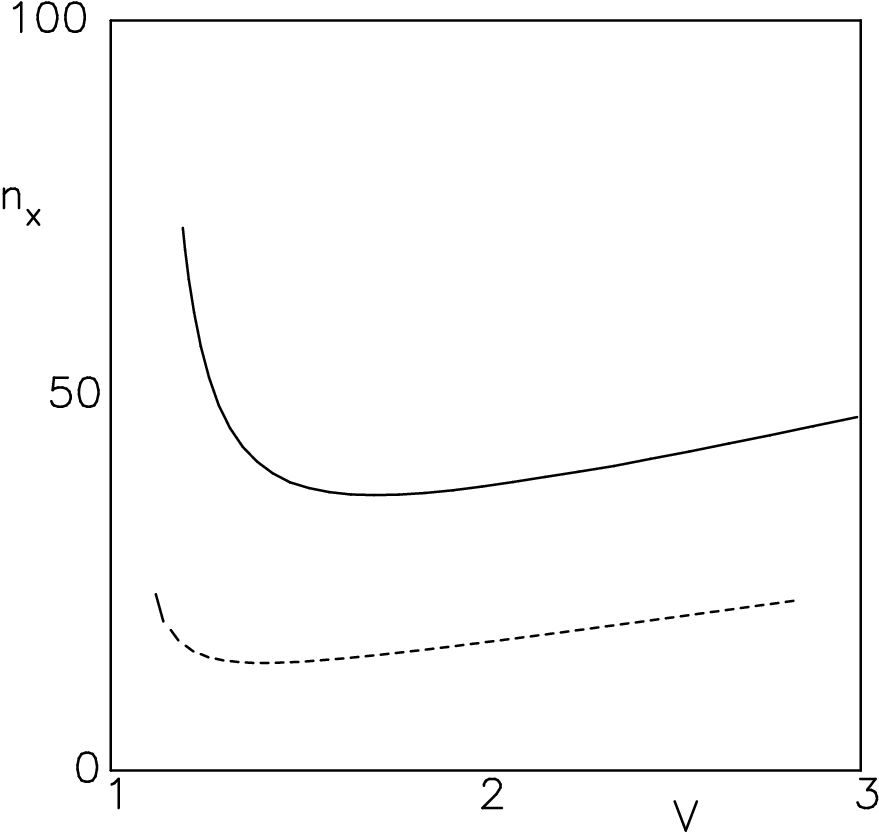}
\caption{How $n_{x}$, the value of particle number $n$ that minimizes
the energy, varies with the strength $V$ of the binding potential.
The interaction parameter was taken to be $g = 2$.  The
dotted line corresponds to the variational result.}
\end{figure}
 
Although the variational results differ quantitatively from the
solution to the GP theory, qualitatively they are similar.  We offer
the following observations: 

*The energy is negative for all $n$, so that the attractive potential has a bound state with an infinite number of particles.

*For $g < g_{c}$, the energy  has a very shallow minimum  
as a function of the number of particles.
This implies that it will be difficult to attach a definite number of particles to a
well by placing it in tunnelling contact with a source (such as a condensate droplet): there
is no equilibrium at all, or the number of particles bound will be large and very susceptible to
small perturbations.
The minimum in $E(n)$ is even less pronounced in the GP
solution than it is in the variational result. 
We have indicated the apparent position of the minima, using the "+" graph
marker.

*For $g > g_{c}$, the energy increases with the number of particles.  Then it is possible to 
bind a well-defined number of particles by letting the system equilibrate with a source.
However, this scheme will only work well for a small number of particles: the plateau at
large $n$ will make the system insensitive to the number of particles.

In the variational study, we found a counterintuitive behavior for
the value $n_{x}$ at which the minimum of the total energy occurs:
when $g$ is below the
critical value $g_{c}$, $n_{x}$ becomes large, diverging as $V$ approaches the binding threshold.
This means that an arbitrarily large number of particles can be
stably bound by making the potential weak!
We have verified that this behavior is also exhibited by the solution to
the GP theory.
Here it is useful to observe that we are trying to find the minimum of the total energy
both with respect to the parameter $n$ and the normalized wavefunction $\psi$.  
Imposing the condition that the derivative of Eq.(1) with respect to
$n$ (with $\psi$ fixed) is zero, and combining the outcome with
Eq. (\ref{relation}), we
obtain a relationship
\begin{equation}
\label{mincond}
g = 2 |\epsilon|/B . 
\end{equation}
Imposing this condition proved to be easier than trying to minimize $E$ over $n$.
We constructed Figure 2 by finding
normalized solutions to the GP equation (\ref{diffGPE})
for various values of $\gamma$, and iterating until
Eq. (\ref{mincond}) holds.

Although our work demonstrates the existence of the effect of resonant 
binding only within the GP approximation, we expect that it
is also present in the exact theory.  As a nearly-resonant well is
made more shallow approaching the single-particle binding threshold,
the effective range of the central attraction increases, and the well
binds an increasingly larger number of bosons.  Therefore the microscopic
details of the well are not expected to be of qualitative
significance.  The large range is
responsible for the survival of the effect in the presence of weak 
short-range repulsion between the bosons because it allows the
particles to take advantage of the central attraction while 
minimizing mutual contact interactions \cite{note}.  The
interactions become more important as the well is made
deeper, thus explaining the existence of the minimum in Fig. 2.  For
a sufficiently deep well the number of bound particles $n_{x}$ increases as the
well deepens which is the normally expected behavior.

Additionally, we argue that our results are immune to the effects of 
correlations neglected in the GP approximation.  Indeed, the GP theory
can be viewed as a variational method.  Therefore the exact energy
will be lower than our calculations indicate.  What they do
show is that near resonance the wavefunction can accommodate an
unexpectedly large number of particles, because each particle manages
to sample the attractive well without interacting with the
other particles too strongly.  The effect of correlations would be
that the particles evade each other more effectively, and thus enhance 
the effect we predict. 

The issue of validity of the GP theory was also addressed in
Ref.\cite{KSK} where the same
problem was studied in general dimension.    Specifically, in one dimension a version of
the problem in question was solved exactly by using the Bethe ansatz
methods and the results were compared with the GP
solution.  Although the exact ground-state wavefunction is indeed not of the
product (Hartree) type, in the limit of large particle number the
exact ground-state energy turns out to be nearly identical to its 
GP counterpart.  This example represents ``the worst case scenario''
as the effect of correlations is expected to be strongest in one dimension.
Therefore we conclude that it is very likely that the GP theory is
fully adequate in three spatial dimensions.          

\section{Weakly attracting particles}

The variational analysis also predicts an interesting phenomenon for the case
of weakly-attracting particles in the presence of a potential that is too weak to
bind a single particle ($g < 0$ and $V \le 1$).
The second term of Eq. (\ref{varEnApp}) is negative for sufficiently large $n$, pointing to the
existence of a many-body bound state.
For the particular case $V=1$, the critical value of the number of particles needed is given by
$n_{c} = 3 \pi/2 |g|$.
When $V < 1$, the minimum energy can be positive (indicating a metastable state), but becomes negative
for sufficiently large $n$.
The possibility of a cooperatively bound state of this sort was first
pointed out  by Migdal \cite{migdal}.  

\begin{figure}
\includegraphics[
  width=1.0\columnwidth,
  keepaspectratio]{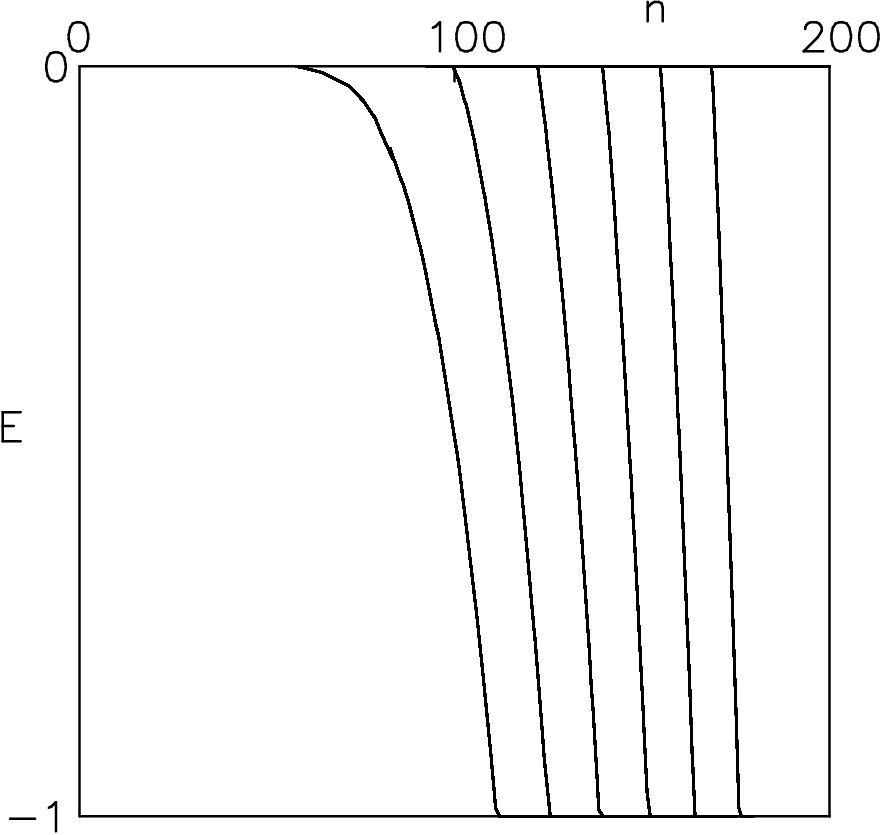}
\caption{The ground-state energy of weakly-attractive bosons as a
function of their number $n$, for various values of the binding
potential amplitude $V$.  The curves are all for the case $g = -0.1$,
and the successive curves from left
to right. are for $V = 1.00,0.98,0.96,0.94,0.92,0.90$.}
\end{figure}

Figure 3 shows the results of numerical solution of the
GP theory for the case of weakly attracting particles in the
presence of a $V \le 1$ potential.   In agreement with the variational
argument, collective bound states are seen to form for a sufficiently 
large number of bosons.  Specifically, the farther away is the binding
threshold, the larger number of bosons is required to form a bound
state. 

\section{Beyond the model}

The delta-shell potential cannot be achieved experimentally.  Then one may ask
how our results are relevant to the real world.

The variational results are quite general for a localized external potential.
Since the trial wavefunction is constant for $r < a$, the external potential
only enters through its average value
\begin{equation}
a^2 V = - \int \limits_{0}^{\infty}U(r)r^{2}dr
\end{equation}
The solution to the GP equation will be affected slightly.  We
can anticipate the effects by comparing the delta shell problem to that of
the attractive square well $U(r) = -W$ for $r < a$.
Outside the
well the GP equation is the same for the square well and the delta shell, and
so the solutions (for given $\gamma$ and $\epsilon$) are the same; the only question
is how the value for $W$ corresponds to that for $V$.
The condition that relates them is that the values of the logarithmic derivative of
the wavefunctions should be the same at $r = a$. 
For the delta shell, the solution to Eq.(\ref{diffGPE}) (for $\gamma = 0$) 
is $\psi = C \sinh(\sqrt {|\epsilon|} r)/r$, and  is very nearly constant for the
cases of interest; however
for the square well the solution inside (again for noninteracting particles)
has the form $\psi = C \sin(\sqrt{W - |\epsilon|}r)/r$.
The logarithmic derivative of this function takes on all values and
there is an infinite set of values of $W$ for which the match can be achieved.
The effect of the particle interactions will be to somewhat suppress the 
accumulation of particles (places where $\psi$ is large) and to 
push the consecutive bound states to larger values of $W$.
 
However, the feature of interest in the resonantly bound system is the
number of particles that can be bound, and this is a property of
the wavefunction at large distances.  Thus we can expect that the 
the differences between different external potentials can be absorbed
into the definitions of $V_{c}$ and $g$.
\vspace{0.5cm} 
\section{Conclusions}

To summarize, our work demonstrates that a super-threshold
localized trapping potential can bind an unexpectedly large number of
weakly-repulsive bosons while a sub-threshold trap can trigger a
formation of a cooperative bound state of weakly-attractive bosons
even when two-body attraction is insufficient to form a two-body
state.  These effects are insensitive to the microscopic details of
the trapping potential and have their origin in the large effective
range of nearly-resonant trapping potential.  We hope that these
results will be experimentally tested in the near future.    

\section{ACKNOWLEDGMENTS}

This work was supported by the Thomas F. Jeffress and Kate 
Miller Jeffress Memorial Trust.

\end{document}